\documentstyle[prl,aps,epsfig,multicol]{revtex}

\onecolumn

\begin{document}
\title{Precision frequency measurements with entangled states}
\author{Christian~F.~Roos\cite{byline}}
\address{Institut f{\"u}r Quantenoptik und Quanteninformation, {\"O}sterreichische Akademie der Wissenschaften,\\
Technikerstra{\ss}e 21a, A-6020 Innsbruck, Austria}
\date{\today}
\maketitle

\begin{abstract}
We demonstrate how quantum entanglement can be used for precision
frequency measurements with trapped ions. In particular, we show
how to suppress linear Zeeman shifts in optical frequency
measurements by using maximally entangled states of two ions even
if the individual ions do not have any field-independent
transition. In addition, this technique allows for an accurate
measurement of small external field frequency shifts such as the
electric quadrupole shift which are important for ion clock
experiments.
\end{abstract}

\pacs{PACS numbers: 06.30.Ft,03.67.-a,32.80.Qk,39.30.+w}




\begin{multicols}{2}

Frequency standards based on narrow-linewidth optical transitions
in neutral atoms and ions hold great promise as future primary
frequency standards\cite{Madej01,Diddams01}. In an optical clock,
a laser interrogating an atomic transition serves as a local
oscillator. The laser frequency is regulated by a servo loop to
match the resonance of the atomic transition; a frequency comb
divides the optical frequency down to the microwave regime where
it can be counted electronically. Laser-cooled single ions with
forbidden transitions connecting the ground to a long-lived
metastable state have the potential of achieving quantum-limited
instabilities of about $10^{-15}$ \cite{Diddams01} for 1s
averaging times and fractional frequency uncertainties down to
$10^{-18}$ \cite{Dehmelt82}. To achieve this exquisite level of
precision, transitions between atomic levels have to be chosen
whose transition frequencies are highly immune to external
magnetic and electric fields. Great care has to be taken to
determine level shifts caused by unavoidable fields (e.~g.
black-body radiation) with high precision.

The capability of deterministically entangling quantum systems is
a key element in many quantum information and communication
protocols. For quantum measurements involving neutral atoms and
ions, techniques developed in this context\cite{Giovannetti04}
allow for improved parameter estimation in interferometric setups,
for efficient quantum state detection\cite{Schaetz05,Schmidt05},
the measurement of scattering lengths\cite{Widera04} and for
beating the shot noise limit. Even prior to the surge of interest
in quantum information processing, the use of quantum-mechanical
correlations for improving optical clocks was being investigated.
It has been shown that N-ion entangled states can lead to a
sensitivity not attainable with the same number of uncorrelated
atoms \cite{Bollinger96,Wineland98} and first demonstration
experiments have been performed\cite{Meyer01,Leibfried04}. This
technique is a generalization of Ramsey spectroscopy to entangled
states. While the gain in measurement sensitivity due to
entanglement between the ions depends on the model under
consideration \cite{Wineland98,Huelga97}, it is always smaller
than $\sqrt{N}$ and becomes considerable only if large numbers of
ions can be reliably prepared in a maximally entangled state.

In this paper, we will discuss a different application of
maximally entangled states for high-precision spectroscopy that
requires only two ions to be entangled. We will show that certain
external-field-dependent level shifts can either be cancelled or
precisely measured by preparing the two-ion maximally entangled
state in a superposition involving several different internal
atomic levels. This leads to two possible applications of
entangled states: (1) the measurement of optical frequencies with
increased immunity to decoherence caused by fluctuating magnetic
fields and (2) the measurement of small frequency shifts caused by
external electric and magnetic fields. Recently, it has been shown
that entanglement is able to persist for up to 1s, limited only by
spontaneous decay of the metastable atomic level\cite{Roos04}.
This clearly demonstrates the potential entangled states have for
precision measurements.

All ions investigated as potential optical frequency
standards\cite{Madej01} have narrow optical transitions connecting
the ground state with a metastable excited state that typically
has a lifetime of several hundred milliseconds or more. Most of
them are either hydrogen-like ions with half-integer nuclear spin
such as $^{199}$Hg$^+$,$^{171}$Yb$^+$,$^{87}$Sr$^+$ or have
alkali-earth-like spectra, e.~g. $^{115}$In$^+$ and $^{27}$Al$^+$.
While these ions have transitions with either zero or very small
first-order Zeeman shift, such favorable transitions do not exist
in even-isotope hydrogen-like ions such as
$^{88}$Sr$^+,^{40}$Ca$^+,^{138}$Ba$^+$ and $^{172}$Yb$^+$. Here,
fluctuating magnetic fields may preclude the observation of the
natural line width of narrow resonance lines from the $S_{1/2}$ to
the $D_{5/2}$ level. Careful magnetic shielding of the apparatus
is required to observe narrow lines. From this point of view, even
isotopes seem to be much less attractive as frequency standards.

Yet, entangled states make it possible to cancel the first-order
Zeeman effect even in those ions that have no first-order-free
transitions. The cancellation method is inspired by the technique
described in ref. \cite{Bollinger96} but makes use of the
multi-level structure of real atoms. For a single two-level ion
with ground state $|g\rangle$ and excited state $|e\rangle$,
separated by the energy $\hbar\omega_0$, the transition frequency
can be measured by Ramsey spectroscopy: a laser pulse with
frequency $\omega_L$ and pulse area $\pi/2$ coherently excites the
ion to the state $(|g\rangle+|e\rangle)/\sqrt{2}$. In a reference
frame rotating with frequency $\omega_L$ this superposition
evolves during a period of free evolution of duration $\tau$ into
the state $(|g\rangle+e^{-i\Delta\tau}|e\rangle)/\sqrt{2}$ where
$\Delta=\omega_0-\omega_L$. A second $\pi/2$ pulse finally maps
this state onto an unequal superposition of $|g\rangle$ and
$|e\rangle$ so that by measuring the population difference
$p_e-p_g=\cos(\Delta\tau)$ information about the difference
between atomic and laser frequency is inferred. The Ramsey method
is applicable to entangled states in the following sense: The
first $\pi/2$ pulse is replaced by a sequence of laser pulses that
prepares the two-ion Bell state
$(|g\rangle|g\rangle+|e\rangle|e\rangle)/\sqrt{2}$. This state
evolves during free precession into
$(|g\rangle|g\rangle+e^{-i2\Delta\tau}|e\rangle|e\rangle)/\sqrt{2}$.
After applying $\pi/2$ pulses to both ions, a measurement of
$p_{ee}+p_{gg}-p_{eg}-p_{ge}=\cos(2\Delta\tau)$ reveals the
deviation of the laser frequency from the atomic resonance. Here,
$p_{ij}$ denotes the joint probability of detecting ion 1 in state
$|i\rangle$ and ion 2 in state $|j\rangle$. By identifying the
state $|e\rangle$ ($|g\rangle$) with the eigenstate of the Pauli
spin operator $\sigma_z$ with positive (negative) eigenvalue, this
measurement gives the expectation value of the spin correlation
$\sigma_z^{(1)}\sigma_z^{(2)}$.

\begin{center}
\begin{figure}[tbp]
\epsfxsize=0.95\hsize \epsfbox{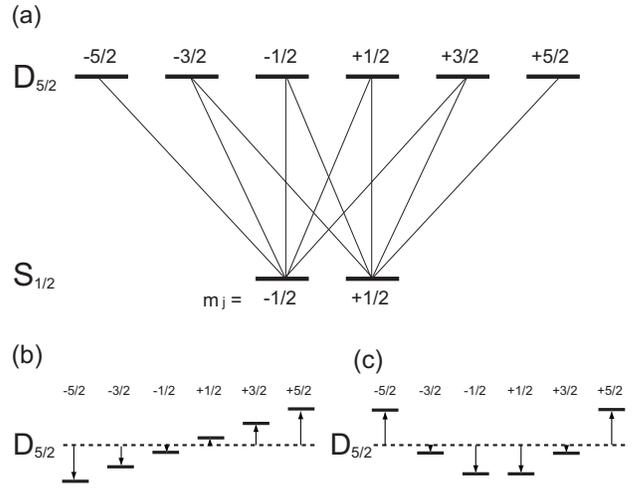} \caption{(a) Zeeman
levels of the $S_{1/2}$ and $D_{5/2}$ states and electric
quadrupole transitions allowed by the selection rules. (b) In a
magnetic field, the energy shift of the Zeeman levels $m_j^\prime$
is $\Delta E_Z=hg_{D_{5/2}} m_j^\prime B$ to first order. (c)
Static electric field gradients couple to the quadrupole moment of
the $D_{5/2}$ state and shift the levels by an amount $\Delta
E_Q\propto 3{m_j^\prime}^2-j(j+1)$. AC electric fields coupling to
the tensor polarizability of the $D_{5/2}$ state lead to level
shifts with the same dependence on $m_{j^\prime}$. Note that under
typical experimental conditions the Zeeman shift is about five
orders of magnitude bigger than the quadrupole shift.\label{fig1}}
\end{figure}
\end{center}

Now, the cancellation of the first order Zeeman effect in even
isotopes is achieved by applying the generalized Ramsey method to
Bell states of the type
\begin{equation}
|\psi_0\rangle=\frac{1}{\sqrt{2}}(|s_+\rangle|s_-\rangle+
e^{i\phi_0}|m_j'\rangle|-\!m_j'\rangle). \label{qstate}
\end{equation}
Here and throughout the paper, the abbreviations
$|m_j^\prime\rangle$ and $|s_\pm\rangle$ denote the states
$|D_{5/2},m_j^\prime\rangle$ and $|S_{1/2},m_j=\pm 1/2\rangle$,
respectively. Fig.~\ref{fig1} shows the electric
quadrupole-allowed transitions and the level shifts caused by
external fields. By associating each level shifting upwards in a
magnetic field with another level shifting downwards by the same
amount, the energy of the combined system becomes independent of
the magnetic field strength in first order. As a consequence, the
corresponding \textquoteleft super-transition\textquoteright
 connecting the two-ion ground state to the two-ion excited state
is not affected by changing magnetic fields. Its frequency can be
probed by applying the measurement scheme described above. For the
state $|\psi_0\rangle$ defined in eq.~(\ref{qstate}), the
expectation value yields
$\langle\sigma_z^{(1)}\sigma_z^{(2)}\rangle=\cos(\frac{\Delta_++\Delta_-}{2}\tau-\phi_0)$,
where $\Delta_\pm$ is the detuning of the laser frequency from the
$|s_\pm\rangle\leftrightarrow |\pm\!m_j^\prime\rangle$ transition
frequencies. An advantage of even isotopes stems from the lack of
hyperfine structure which makes the cross-over from the Zeeman to
the Paschen-Back regime occur at much higher magnetic fields,
resulting in a much smaller second-order Zeeman effect. Therefore,
for an accurate calculation of the transition frequency, the
Land\'e factor $g_j(D)$ of the excited level does not need to be
known to the same precision as for odd isotopes.

Frequency measurements with entangled states require a reliable
and deterministic method of producing the entangled initial state.
Ion-trap experiments dedicated to quantum information
processing\cite{Wineland98,SchmidtKaler03} use strings of ions
that are confined in traps providing motional frequencies of a few
MHz and giving rise to inter-ion distances of a few $\mu$m. After
laser cooling one or several normal vibrational modes of the ion
crystal close to the ground state, quantum correlations between
the internal atomic states are generated by coupling the internal
atomic states to their motional degrees of freedom through
laser-atom interactions\cite{CiracZoller95,Wineland98,Molmer99}.
This allows for entanglement generation mediated by the motional
degrees of freedom. For the generation of ion entanglement,
quantum gates or specially tailored series of laser pulses have
been used \cite{Turchette98,Roos04}. In a non-zero magnetic field,
Zeeman transitions can be selectively excited with a narrow-band
laser. Generation of the entangled state $|\psi_0\rangle$ is
straightforward through the individual addressing of ions with a
focused laser beam\cite{SchmidtKaler03}. Individual addressing by
focused laser beams is not, however, absolutely required.
Alternatively, since different laser frequencies are needed to
excite ions 1 and 2, one could also work with an unfocused beam
starting from the state $|s_+\rangle|s_-\rangle$ that could be
prepared by an optical pumping beam in a lin$\perp$lin standing
wave. Starting from $|s_+\rangle|s_-\rangle$, the state
$|\psi_0\rangle$ defined in eq. (\ref{qstate}) could be generated
by applying a blue sideband $\pi/2$ pulse\cite{Wineland98} on the
$|s_+\rangle\leftrightarrow |m_j^\prime\rangle$ transition
followed by a red sideband $\pi$ pulse on the
$|s_-\rangle\leftrightarrow |-\!m_j^\prime\rangle$ transition.

Are there potential sources of errors that would not occur in
frequency measurements with single ions? Slow drifts of control
parameters like laser intensity and trap voltage could lead to
drifts of the relative phase $\phi_0$ of the initial entangled
state causing erroneous frequency measurement results. This type
of error as well as errors arising from imperfect analysis pulses
can be counteracted by setting $e^{i\phi_0}=i$ and by alternating
between measurements with long free precession time $\tau$ and
measurements with $\tau$ set to zero. In this way, only changes of
the relative phase that occur during the free precession enter the
measurement result. The additional measurement will not
significantly lengthen the total measurement time since state
preparation and detection will typically require a time much
shorter than $\tau$.

Also, external field shifts caused by electric fields have to be
considered \cite{Itano00}. For two-ion experiments, the use of a
linear Paul trap would be mandatory to avoid excess micromotion
that would give rise to second-order Doppler shifts and ac-Stark
shifts. In contrast to spherical Paul traps, the linear trap's
static electric potential $\Phi=Q_{dc}(2z^2-x^2-y^2)$, providing
the confinement in the axial $z$ direction, produces an electric
field gradient $dE_z/dz=-m\omega_z^2/q$ where $\omega_z$ is the
center-of-mass oscillation frequency of the ion string in the
axial direction and $m$ and $q$ are the mass and the charge of the
ion, respectively. The interaction of the trap's rotationally
symmetric field gradient with the quadrupole moment of the
$D_{5/2}$ level shifts the Zeeman levels by an
amount\cite{Itano00}
\begin{equation}
\Delta\nu=\frac{dE_z}{dz}\Theta(D,5/2)\frac{(j(j+1)-3{m_j^\prime}^2)}{j(2j-1)}\frac{(3\cos^2\beta-1)}{4h}.
\end{equation}
Here, $\Theta(\gamma,j)=\langle\gamma jj|\Theta_0^{(2)}|\gamma
jj\rangle$ with $\Theta_0^{(2)}=-e/2(3z^2-r^2)$ denotes the
quadrupole moment and $\beta$ the angle between the quantization
axis and the principal axis of the trap. For a two-ion crystal,
the Coulomb potential of the neighboring ion located at a distance
$d=(2q^2/(4\pi\epsilon_0 m\omega_z^2))^{1/3}$ adds an electric
field gradient to the contribution of the trapping potential so
that $(dE_z/dz)_{total}=-2m\omega_z^2/q$ at the equilibrium
positions of the ions. However, since any Zeeman state of the
$D_{5/2}$ level can be used and
$\sum_{m_j^\prime}j(j+1)-3{m_j^\prime}^2=0$, it is possible to
cancel the electric quadrupole shift \cite{Dube05,Margolis04} by
averaging the measurement results obtained with different $m_j'$.
Yet another effect needs to be considered: the phase evolution of
state $|\psi_0\rangle$ is magnetic field-independent only if the
field at the positions of ions 1 and 2 is the same. A residual
magnetic field gradient in the direction of the ions will change
the phase evolution rate. Again, this effect can be averaged out
by measuring the phase evolution for state $|\psi_0\rangle$ and
for a state with the roles of ion 1 and ion 2
interchanged\cite{gradient}, i.~e.
$|\tilde{\psi_0}\rangle=\frac{1}{\sqrt{2}}(|s_-\rangle|s_+\rangle+
e^{i\phi_0}|-\!m_j'\rangle|m_j'\rangle)$.

A second application of entangled states for high-precision
spectroscopy does not even require the ultra-stable lasers needed
for measurements of optical frequencies. Small level shifts can be
precisely determined by preparing Bell states that are immune
against phase decoherence caused by a fluctuating laser frequency.
This measurement technique should allow for a precise
determination of the quadrupole moment of the $D_{5/2}$ level. As
can be seen from Fig.~\ref{fig1}, the state
\begin{equation}
|\psi_1\rangle=\frac{1}{\sqrt{2}}(|5/2\rangle|\!-\!5/2\rangle+|1/2\rangle|\!-\!1/2\rangle)
\label{quadstate}
\end{equation}
is insensitive to the first-order Zeeman shift but evolves in time
in the presence of an electric field gradient as
$|\psi_1(\tau)\rangle=(|5/2\rangle|\!-\!5/2\rangle + e^{i\alpha_1
\tau}|1/2\rangle|\!-\!1/2\rangle)\sqrt{2}$ where
$\alpha_1/(2\pi)=\frac{36}{5}\Delta\nu$ and $\Delta\nu$ is the
quadrupole shift experienced by a single ion in state
$|5/2\rangle$. The increase in sensitivity as compared to the
single-ion state stems from the two-ion entanglement and the fact
that Zeeman levels are chosen that are shifted in opposite
directions by the field-gradient which itself is two times higher
than in the single-ion experiment. If there is a residual magnetic
field gradient in the direction of the ion string, then
$\alpha_1=\alpha_{QS}+\alpha_{B'}$ where $\alpha_{QS}$ denotes the
contribution of the quadrupole shift and $\alpha_{B'}$ the
contribution of the field gradient. The two quantities can be
separated by preparing the state
$|\psi_2\rangle=(|\!-\!5/2\rangle|5/2\rangle+|\!-\!1/2\rangle|1/2\rangle)/\sqrt{2}$
where the roles of ion 1 and ion 2 have been interchanged. This
state will evolve with phase evolution rate
$\alpha_2=\alpha_{QS}-\alpha_{B'}$ provided that
$|\alpha_{B'}|<|\alpha_{QS}|$ so that the quadrupole shift can be
measured by taking $\alpha_{QS}=(\alpha_2+\alpha_1)/2$. Since
patch potential effects could produce an additional field
gradient, a plot of the quadrupole shift versus $\omega_z^2$ will
yield a hyperbola; the quadrupole moment of the $D_{5/2}$ can be
determined from the slope of the tangent to
$\alpha_{QS}(\omega_z^2)$.

D-state quadrupole moments of $^{199}$Hg$^+$,$^{171}$Yb$^+$ and
$^{88}$Sr$^+$ have been recently measured with an accuracy ranging
from $3\%$ to $50\%$ \cite{Oskay05,Barwood04,Schneider05} by
detecting Hz-level changes on top of the transition frequencies at
$\approx 10^{15}$~Hz Using entangled states, a relative
measurement uncertainty of below $1\%$ is to be expected. For
example, in the case of two $^{88}$Sr$^+$ ions stored in a linear
trap with axial frequency $\omega_z=(2\pi)\,1$~MHz, the ions
experience a field gradient $dE_z/dz= 72$~V/mm$^2$. The phase of
$|\Psi_1\rangle$ in eq.~(\ref{quadstate}) would evolve with a
frequency of $152$~Hz if the direction of the magnetic field was
aligned with the trap axis ($\beta=0$) and using the measured
value $\Theta(D,5/2)=2.6\,ea_0^2$. The entangled state decays
within the time $\hat{\tau}=\tau_{D_{5/2}}/2$ where
$\tau_{D_{5/2}}\approx 350$~ms is the lifetime of the $D_{5/2}$
level. This sets an upper limit to the maximum useful free
precession time $\tau$. Therefore, if the relative phase
$\phi_1(\tau)=\alpha_1\tau$ of $|\Psi_1(\tau)\rangle$ after a time
$\tau=150$~ms could be determined with a precision of $0.1$
radian, the phase evolution frequency could be measured with an
uncertainty of $0.2\%$. Then the uncertainty in the measured
quadrupole moment would probably be dominated by the measurement
of the angle $\beta$. If the quantization axis could be aligned
with the trap's z-axis to within $5^\circ$, an uncertainty of
below $1\%$ could be achieved.

Employing entangled states for measurements of small energy shifts
is not limited to the case of quadrupole shifts: (1) The same
measurement principle could also be applied for the determination
of the tensor polarizability of the $D$ levels. For this, a
transverse static field would have to be applied to shift the ion
crystal to a position at which the ions experience a quadratic
Stark effect as demonstrated in \cite{Schneider05,Yu94}. (2)
Another application concerns the measurements of isotope shifts.
By loading a crystal consisting of different isotopes, the isotope
shift on the quadrupole transitions could be measured with
unprecedented precision. (3) Segmented ion traps could be used to
separate an ion Bell pair\cite{Barrett04}. Then, one ion could be
brought close to a surface, thus acting as a miniature sensor of
local electromagnetic fields that is read out by the detection of
the frequency shift experienced by one ion with respect to the
other one. (4) Finally, the search for a possible change of
fundamental constants \cite{Peik04} using trapped ions might
profit from the use of Bell states. For two ions with similar
transition frequencies $\nu_1,\nu_2$ but a different dependence on
the fundamental constant in question, the measurement sensitivity
could be increased by measuring $|\nu_1-\nu_2|$ in a Bell state
instead of detecting $\nu_1$ and $\nu_2$ separately, provided that
$|\nu_1-\nu_2|\ll \nu_1$. The development of frequency combs makes
phase locking of the lasers exciting ion 1 and 2 feasible
\cite{Cundiff03}.

In summary, we have shown how Bell pairs of ions can be used in
precision spectroscopy for absolute frequency and frequency
difference measurements. For atomic clocks, entangled states
prepared in decoherence-free subspaces might make frequency
measurements more robust against environmental perturbations. The
use of entanglement for frequency difference measurements might
facilitate the detection of tiny frequency shifts on top of the
huge frequencies of optical transitions.

I would like to thank P.~Schmidt and H.~H\"affner for useful
discussions and a critical reading of the manuscript. This work
was supported by the Austrian Science Foundation, European
Networks and the Institute for Quantum Optics and Quantum
Information.

\end{multicols}


\begin{references}

\bibitem[*]{byline} Electronic address: Christian.Roos@oeaw.ac.at

\bibitem{Madej01} A.~A.~Madej, and J.~E.~Bernard, in {\sl Frequency
Measurement and Control}, Topics. Appl. Phys. {\bf 79}, 153
(2001).

\bibitem{Diddams01} S.~A.~Diddams {\sl et al.}, Science {\bf 293}, 825
(2001).

\bibitem{Dehmelt82} H.~G.~Dehmelt, IEEE~Trans.~Instrum.~Meas. {\bf
31}, 83 (1982).

\bibitem{Giovannetti04} V.~Giovannetti, S.~Loyd, and L.~Maccone,
Science {\bf 306}, 1330 (2004).

\bibitem{Schaetz05} T.~Schaetz {\sl et al.}, Phys.~Rev.~Lett. {\bf 94},
010501 (2005).

\bibitem{Schmidt05} P.~O.~Schmidt, T.~Rosenband, C.~Langer,
W.~M.~Itano, J.~C.~Bergquist, and D.~J.~Wineland, Science {\bf
309}, 749 (2005).

\bibitem{Widera04} A.~Widera, O.~Mandel, M.~Greiner, S.~Kreim, T.~W.~H{\"a}nsch,
and I.~Bloch, Phys.~Rev.~Lett. {\bf 92}, 160406 (2004).

\bibitem{Bollinger96} J.~J.~Bollinger, W.~M.~Itano, D.~J.~Wineland, and D.~J.~Heinzen,
Phys. Rev. A {\bf 54}, R4649 (1996).

\bibitem{Wineland98} D.~J.~Wineland, C.~Monroe,
W.~M.~Itano, D.~Leibfried, B.~E.~King, and D.~M.~Meekhof, J.~Res.
Natl.~Inst.~Stand.~Technol. {\bf 103}, 259 (1998).

\bibitem{Meyer01} V.~Meyer {\sl et al.}, Phys.~Rev.~Lett. {\bf 86}, 5870 (2001).

\bibitem{Leibfried04} D.~Leibfried {\sl et al.}, Science {\bf 304}, 1476
(2004).

\bibitem{Huelga97} S.~F.~Huelga, C.~Macchiavello, T.~Pellizzari, and
A.~K.~Ekert, M.~B.~Plenio, and J.~I.~Cirac, Phys.~Rev.~Lett. {\bf
79}, 3865 (1997).

\bibitem{Roos04} C.~F.~Roos {\sl et al.}, Phys.~Rev.~Lett. {\bf 92}, 220402
(2004).

\bibitem{SchmidtKaler03} F.~Schmidt-Kaler {\sl et al.}, Appl.~Phys.~B {\bf 77}, 789
(2003).

\bibitem{CiracZoller95} J.~I.~Cirac and P.~Zoller, Phys.~Rev.~Lett. {\bf 74}, 4091
(1995).

\bibitem{Molmer99} K.~M{\o}lmer and A.~S{\o}rensen, Phys.~Rev.~Lett. {\bf 82},
1835 (1999).

\bibitem{Turchette98} Q.~A.~Turchette {\sl et al.}, Phys.~Rev.~Lett. {\bf 81}, 3631 (1998).

\bibitem{Itano00} W.~M.~Itano, J.~Res.~Natl.~Inst.~Stand.~Technol. {\bf 105}, 829
(2000).

\bibitem{Dube05} P. Dub{\'e}, A.~A.~Madej, J.~E.~Bernard, L.~Marmet, J.-S.~Boulanger, and S.~Cundy,
Phys.~Rev.~Lett. {\bf 95}, 033001 (2005).

\bibitem{Margolis04} H.~S.~Margolis {\sl et al.}, Science {\bf 306}, 1355 (2004)

\bibitem{gradient} Alternatively, the magnetic field gradient
could be precisely measured by monitoring the phase evolution of
the state
$(|s_+\rangle|s_-\rangle+|s_-\rangle|s_+\rangle)/\sqrt{2}$.

\bibitem{Oskay05} W.~H.~Oskay, W.~M.~Itano, and J.~C.~Bergquist, Phys.~Rev.~Lett. {\bf 94}, 163001
(2005).

\bibitem{Barwood04} G.~P.~Barwood, H.~S.~Margolis, G.~Huang, P.~Gill, and H.~A.~Klein, Phys.~Rev.~Lett. {\bf 93}, 133001
(2004).

\bibitem{Schneider05} T.~Schneider, E.~Peik, and Chr.~Tamm, Phys.~Rev.~Lett. {\bf 94}, 230801
(2005).

\bibitem{Yu94} N.~Yu, X.~Zhao, H.~Dehmelt, and W.~Nagourney, Phys.~Rev.~A {\bf 50},
2738 (1994).

\bibitem{Barrett04} M.~D.~Barrett {\sl et al.}, Nature {\bf 429}, 737
(2004).

 \bibitem{Peik04} E.~Peik, B.~Lipphardt, H.~Schnatz, T.~Schneider, and Chr.~Tamm, and S.~G.~Karshenboim,
  Phys.~Rev.~Lett. {\bf 93}, 170801 (2004).

\bibitem{Cundiff03} S.~T.~Cundiff, and J.~Ye, Rev.~Mod.~Phys. {\bf 75}, 325
(2003).

\end{references}
\end{document}